\documentclass[twoside]{article}
\usepackage{fleqn,espcrc2}
\usepackage{graphicx}
\usepackage{here}

\newcommand{\AmS}{{\protect\the\textfont2
    A\kern-.1667em\lower.5ex\hbox{M}\kern-.125emS}}										
\def\beq{\begin{equation}}
\def\eeq{\end{equation}}
\def\bea{\begin{eqnarray}}
\def\eea{\end{eqnarray}}
\def\bq{\begin{quote}}
\def\eq{\end{quote}}

\def\nnb{\nonumber}
\def\ga{\left(}
\def\dr{\right)}

\def\lrar{\Longrightarrow}
																
\def\nnb{\nonumber}
\def\la{\langle}
\def\ra{\rangle}
\def\nin{\noindent}
\def\ba{\vspace*{-0.2cm}\begin{array}}
\def\ea{\end{array}\vspace*{-0.2cm}}

\def\b{$\bullet~$}
\def\als{\alpha_s}

\def\gg2{ \la\alpha_s G^2 \ra}
\def\gg3{g^3f_{abc}\la G^aG^bG^c \ra}
\def\ggg4{\la\als^2G^4\ra}

\def\calD{ {\cal D} }
\def\ftilde{\tilde f}

\def\therho{\theta\rho}

\newcommand{\pipi}{\mbox{$\pi\pi$}}

\def\frac#1#2{{#1\over#2}}

\def\calD{ {\cal D} }
\def\ftilde{\tilde f}

\def\therho{\theta\rho}
\def\sRa{s_{Ra}}
\def\sRb{s_{Rb}}
\def\fpia{f_{\pi a}}
\def\fpib{f_{\pi b}}
\def\fKa{f_{K a}}
\def\fKb{f_{K b}}
\def\ftildePiaa{\tilde f_{\pi aa}}
\def\ftildePiab{\tilde f_{\pi ab}}
\def\ftildePibb{\tilde f_{\pi bb}}
\def\ftildeKaa{\tilde f_{K aa}}
\def\ftildeKab{\tilde f_{K ab}}
\def\ftildeKbb{\tilde f_{K bb}}

\def\gpia{g_{\pi a}}
\def\gpib{g_{\pi b}}
\def\gKa{g_{K a}}
\def\gKb{g_{K b}}

\title
{\bf{\boldmath
{\Large 
The $\sigma$  and $f_0(980)$ from  $ Ke4$ $\oplus$  $\pi\pi$ scatterings data
} }}

\author{
G. Mennessier \address {\footnotesize Laboratoire
de Physique Th\'eorique et Astroparticules, IN2P3-CNRS \& Universit\'e
de Montpellier II, Case 070, Place Eug\`ene
Bataillon, 34095 - Montpellier Cedex 05,
France.}\,\thanks{{\it E-mail addresses:} gerard.mennessier@lpta.univ-montp2.fr (G. Mennessier), snarison@yahoo.fr (S. Narison), wangxuangong@pku.edu.cn (X.-G. Wang).}
, S. Narison\,$^{\rm{a}}$\,\thanks{Corresponding author.} ,
X.-G. Wang\,$^{\rm{a}}$\,\thanks{China scholarship council fellow under contract n$^0$ 2009601139.}\,\address {\footnotesize Department of Physics, Peking University, Beijing 100871, China.}
}

\begin{document}

\pagestyle{myheadings}
\markright{ }
\begin{abstract}
\noindent
We systematically reconsider, within an improved ``analytic $K$-matrix model", the extraction of the $\sigma\equiv f_0(600)$ and $f_0(980)$ masses, widths and hadronic couplings using new $Ke4\equiv K\to\pi\pi e\nu_e$ data on $\pi\pi$ phase shift below 390 MeV and different sets of $\pi\pi\to \pi\pi / K\bar K$ scatterings data from 400 MeV to 1.4 GeV. 
Our results are summarized in Tables \ref{tab:elastic}, \ref{tab:sigma} and \ref{tab:param}.  In units of MeV, the complex poles are: $M_\sigma=452(12)-{\rm i}~260(15)$
and $M_f=981(34)-{\rm i}~18(11)$, which are comparable with some recent high-precision determinations and with PDG values. 
Besides some other results, we find: $|g_{\sigma K^+K^-}|/|g_{\sigma\pi^+\pi^-}|=0.37(6)$ which confirms a sizeable  $g_{\sigma K^+K^-}$ coupling found earlier, and which disfavours a large $\pi\pi$ molecule or four-quark component of the $\sigma$, while its broad $\pi\pi$ width (relative to the one of the $\rho$-meson) cannot be explained within a $\bar qq$ scenario. The narrow $\pi\pi$ width of the  $f_0(980)$ 
and the large value: $|g_{fK^+K^-}|/|g_{f\pi^+\pi^-}|=2.59(1.34)$, excludes its pure $(\bar uu+\bar dd)$ content.  A significant gluonium component eventually mixed with $\bar qq$ appears to be necessary for evading the previous difficulties. 
\end{abstract}
\maketitle
%%%%%%%%%%%%
\vspace*{-1.5cm}
\section{Introduction}
\vspace*{-0.25cm}
 \nin
%%%%%%%%%%%%
Understanding the nature of scalar mesons in terms of quark and gluon
constituents is a long standing puzzle in QCD  \cite{MONTANET,SN09}.
The problem here is that some  states are very broad
($\sigma$ and $\kappa$\,\cite{MOUSSALAM} (if confirmed) mesons) and others are close to an inelastic
 threshold ($f_0(980)$, $a_0(980)$), which makes their interpretation
 difficult. 
 Besides the interpretation within a $q\bar q$ model
\cite{MONTANET,MORGAN,BN,SNG,KLEMPT,OCHS,SN06}
% \cite{MONTANET,OCHS,BN,SNG,SN06} 
or unitarized quark model \cite{TORN,BEVEREN},
 also the possibility of
tetraquark states \cite{JAFFE1,BLACK,LATORRE,SNA0,ACHASOV} (and some other related
 scenarios: meson-meson molecules \cite{ISGUR,BARNES}, 
%unitarized quark model \cite{TORN}, 
meson exchange \cite{HOLINDE}) is considered. 
In addition, a gluonic meson is expected back to 1975 \cite{MIN} as a consequence of the QCD confinement.  An indication of the existence of scalar glueballs \cite{SN09,VENTO}
comes from phenomenological studies \cite{MINK,OLLER,CHANOW}, lattice QCD \cite{PEARDON,MICHAEL}, strong coupling approach in a Nambu-Jona-Lasinio like-model \cite{FRASCA}, QCD spectral sum rules (QSSR) 
\cite{SNG,SN06,NSVZ,CHET,SNG0,SNG1,VENEZIA,} \`a la SVZ \cite{SVZ,SNB},
some low-energy theorems (LET) \cite{VENEZIA,NOVIKOV,CHANO},  AdS/QCD \cite{HERY} and large $N_c$ \cite{PELAEZ,ARRIOLA}. Such a
state could mix with the other $\bar qq$  mesons 
 \cite{BN,SNG,OCHS,CLOSE,aniso}. 
Among the light particles, the 
$\sigma\equiv f_0(600)$  (hereafter called $\sigma$) meson could be such a gluonic 
 resonance. The $\sigma$ can manifest itself in some effective linear sigma models
 \cite{LANIK,SCHEC,ZHENG} or contribute to the low-energy constants at ${\cal O}(p^4)$ of
the QCD effective chiral Lagrangian \cite{PICH}, while its r\^ole in nuclear matter (e.g. nuclear potential) from its coupling
to nucleons is essential \cite{CHANFRAY}. \\
 One might expect that
the hadronic and two photons couplings of these mesons could provide an important information about their intrinsic composite structure. Indeed, recent analyses of  $\gamma\gamma \to \pi\pi$ data \cite{MNO} and of $\pi\pi \to \pi\pi/\bar KK$ \cite{KMN} indicate that the $\sigma$ meson could be such a gluonic  
 resonance (gluonium/glueball).\\
 In this letter, we pursue the test of the nature of the $\sigma$ and of the $f_0(980)$ by reconsidering the extraction of their parameters (masses and couplings) from  $\pi^+\pi^-$ scatterings 
 using recent  precise data from NA48/2 on $Ke4\equiv K\to\pi\pi e\nu_e$ for the $\pi\pi$-phase shift below 390 MeV \cite{NA48} and different sets of $\pi\pi\to\pi\pi/\bar KK$ data above 400 MeV\,\cite{MUNICH,HYAMS,COHEN,ETKIN,KLL} . 
  %%%%%%%%%%%%%%%%%%%%%%%%
%\vspace*{-0.5cm}
\section{The analytic K-matrix model for $\pi\pi\to\pi\pi/\bar KK$}
\vspace*{-0.25cm}
\nin
%%%%%%%%%%%%%%%%%%%%%%%%
In  this approach, the strong processes are described by a K
matrix model representing the amplitudes by a set of resonance poles\,\cite{MENES}\,\footnote{Some applications of the model have been discussed in \cite{PEAN,LAYSSAC}.} .
In that case, the dispersion relations in the multi-channel case 
can be solved explicitly, which is not possible otherwise. 
The  model can be reproduced by a set of Feynman diagrams, including 
resonance (bare) couplings to  $\pi\pi$ and $K\bar K$
 and (in the original model \cite{MENES}) 4-point $\pi\pi$ and $K\bar K$ interaction vertices
 which we shall omit for simplicity in \cite{MNO} and here.  
A subclass of bubble pion
 loop diagrams including resonance poles in the s-channel are resummed
 (unitarized Born).  In this letter, we discuss the approach for the case of : 
 {\it 1~channel $\oplus$ 1~``bare" resonance (K-matrix pole)} and
{\it 2~channels $\oplus$ 2~``bare" resonances} 
and we restrict to the $SU(3)$ symmetric shape function. \\ In the present analysis, the introduction of a real analytic form factor {\it shape function}, which takes
explicitly into account left-handed cut singularities for the strong interaction amplitude,
allows a more flexible parametrisation of the $\pi \pi\to\pi\pi/\bar KK$ data. 
In our low energy approach, it can be conveniently
approximated by:
  \beq
f_P(s)=\frac{s-s_{AP}}{s+\sigma_{DP}} \label{formfactor}~,~~P\equiv \pi,~K~,
\eeq    
which multiplies the scalar meson couplings to $\pi\pi/\bar KK$. In this form, the {\it shape function} allows for an Adler  zero at $s=s_{AP}$
and a pole at $\sigma_{DP}>0$ simulating the left hand cut. 
%%%%%%%%%%%%%%%%%%%%%%%%%%%
\\
\b {\it \bf 1~channel $\oplus$ 1~``bare" resonance}\\
%%%%%%%%%%%%%%%%%%%%%%%%%%%
Let's first illustrate the method in this simple case. 
The unitary $PP$ amplitude is then written as:
\beq
  T_{PP}(s) = \frac{G_P f_P(s)}{s_R-s -  G_P \ftilde_P (s)} = \frac{G_P f_P(s)}{\calD_P(s)}~,
\label{tpipi}
\eeq 
where $T_{PP}=e^{i\delta_P}\sin\delta_P/\rho_P(s)$ with 
 $ \rho_P(s) =({1 - 4 m^2_P/s})^{1/2}$;  $G_P=g_{P\sigma,B}^2$
 are the bare coupling squared and :
\beq
{\rm {Im}}~ \calD_P = {\rm{Im}} ~ (- G_P \ftilde_P) = - (\therho_P) G_P \ f_P~,  
\label{eq:imaginary}
\eeq
with: $(\theta\rho_P)(s)=0$ below and $(\theta\rho_P)(s)=\rho_P(s)$
above threshold $s=4m_P^2$. The ``physical" couplings are defined from the residues,
with the normalization:
\beq
g_{P\sigma}^2\equiv g^2_{\sigma PP}/(16\pi)~.
\eeq
The amplitude near the pole $s_0$ where $ {\cal D}_P(s_0)=0$ and
${\cal D}_P(s)\approx {\cal D'}_P(s_0) (s-s_0)$ is:
\beq
  T_{PP}(s)\sim \frac{g_P^2}{s_0-s}; \qquad g_P^2=\frac{G_P
f_P(s_0)}{-\calD'(s_0)}~.
\label{eq:gpi2}  
\eeq
The real part of $\calD_P$ is obtained from a dispersion relation with
subtraction at $s=0$ and one obtains:
\beq
\ftilde_P(s) = \frac{2}{\pi} \Big{[} h_P(s) \ -h_P(0) \Big{]}~:
\label{eq:ftilde}
\eeq
$h_P(s) =f_P(s) \tilde L_{s1}(s)$--$(\sigma_{NP}/(s+\sigma_{DP}))\tilde L_{s1}(-\sigma_{DP})$, 
$\sigma_{NP}$ is the residue of $f_P(s)$ at $-\sigma_{DP}$ and: $\tilde L_{s1}(s) =  
 \big{[}\ga{s - 4 m_P^2}\dr/{m_P^2} \big{]}
\tilde L_1(s,m_P^2)$ with $\tilde L_1$ from \cite{MENES}. \\
%%%%%%%%%%%%%%%%%%%%%%%%%%%
\b {\bf 2 channels $\oplus$ 2 ``bare" resonances} \\
%%%%%%%%%%%%%%%%%%%%%%%%%%%
The generalization to this case is conceptually straightforward though cumbersome. 
Let us consider two 2-body channels coupled to 2 ``bare" resonances  labelled $a$ and $b$,
with bare masses squared  $s_{Ra}$ and $s_{Rb}$:\\
-- Let $\fpia(s)$ , $\fpib(s)$ ,$\fKa(s)$ , $\fKb(s)$ be four {\it shape functions},
real analytic in the s-plane, with {\it left cut},
and $\ftildePiaa (s)$, $\ftildePibb (s)$ , $\ftildePiab (s)$,
    $\ftildeKaa (s)$,  $\ftildeKbb (s)$ ,  $\ftildeKab (s) $,
six functions, real analytic in the s-plane, with {\it right cut}. Their imaginary parts on the cut for $s \ge 4 m_P^2$ are:
\bea
  {\rm Im} \ftildePiaa (s+i\epsilon) &=& (\therho_\pi \ \fpia ^2)(s)
                                ~, \\
                 {\rm Im} \ftildePibb (s+i\epsilon) &=& (\therho_\pi \ \fpib ^2)(s)
                                ~, \nnb\\
                  {\rm Im} \ftildePiab (s+i\epsilon) &=& (\therho_\pi \ \fpia \ \fpib)(s)
                                ~,\nnb
\eea
and analogous for the 2nd $\bar KK$ channel. \\
-- Let's define the bare inverse propagators:
\beq
 D_a(s) = (\sRa - s);  \quad  D_b(s) = (\sRb - s)~,
 \eeq 
 and the ``bare" couplings $\gpia$, $\gpib$, $\gKa$, $\gKb$ of the resonances to the channels,
through the pure 1-resonance inverse propagators:
\bea 
\calD_a(s) &=& D_a (s)- \gpia^2 \ \ftildePiaa (s)- \gKa^2 \ \ftildeKaa(s)~,\nnb\\
  \calD_b(s) &=& D_b (s)- \gpib^2 \ \ftildePibb(s) - \gKb^2 \ \ftildeKbb(s)~.
\eea
-- Let's define
the full denominator function $\calD(s)$, analytic in the s-plane,
with {\it right cut} $s \ge 4 m_\pi^2$ :
\bea
  \calD (s)
    &=& \calD_a \ \calD_b
      - (\gpia \gpib \ftildePiab  + \gKa \gKb \ftildeKab  )^2  \nnb\\
&=& D_a D_b - D_a \ ( \gpib^2 \ \ftildePibb  + \gKb^2 \ \ftildeKbb  )\nnb\\
     && - D_b \ ( \gpia^2 \ \ftildePiaa  
    + \gKa^2 \ \ftildeKaa  )  
      + (\gpib^2 \ \ftildePibb  \nnb\\
      &&+ \gKb^2 \ \ftildeKbb  )
                          (\gpia^2 \ \ftildePiaa  
                          + \gKa^2 \ \ftildeKaa  )\nnb\\
    && - ( \gpia \gpib \ftildePiab  + \gKa \gKb \ftildeKab  )^2~,                                                                  
\eea
and the partial propagators
\bea
  P_{aa} &=& \frac{\calD_a}{\calD}~,~ P_{bb} = \frac{\calD_b}{\calD}~,\nnb\\
  P_{ab} &=&\frac{1}{\calD}( \gpia \ \gpib  \ftildePiab + \gKa \ \gKb  \ftildeKab)~.
  \eea
-- Then
\bea
  T_{\pi \pi}
            &=& \gpia^2 \fpia^2          P_{aa} 
          + 2 \gpia \gpib \fpia \ \fpib P_{ab} 
            + \gpib^2 \fpib^2         P_{bb},
                     \nnb\\  
T_{K K}&\equiv &T_{\pi \pi}~: ~\pi \to K,  
            \nnb\\                    
  T_{\pi K} &=& T_{K \pi}\nnb\\
  &=&  \gpia \gKa \fpia \ \fKa                         P_{aa} 
            + (\gpia \gKb \ \fpia \fKb \nnb\\
            &&+ \gKa\gpib \ \fKa \fpib)  P_{ab} 
            + \gpib \gKb \fpib \ \fKb                          P_{bb},\nnb\\                   
\eea
is a set of unitary elastic amplitudes. \\
-- The inelasticity $\eta$ is related to the amplitudes or $S$-matrix as:
\bea
&&\eta e^{2i\delta_P}=S_{PP}\equiv 1+2~{\rm i} \rho_PT_{PP}~,~P\equiv \pi,~K~,\nnb\\
&&\sqrt{1-\eta^2} e^{i\delta_{\pi K}}\equiv -{\rm i}~S_{\pi K}=2 \sqrt{\rho_\pi\rho_K}T_{\pi K}~,
\eea
where the sum of pion and kaon phase shifts is:
\beq
\delta_{\pi K}=\delta_{\pi}+\delta_{K}~.
\eeq
-- In the following, we shall work in the {\it minimal case} with one shape function:
\beq
\fpia(s)=\fpib(s)=\fKa(s)=\fKb(s)~,
\eeq
where:
\beq
\sigma_D\equiv \sigma_{D\pi}=\sigma_{DK}~;~~~~
s_A\equiv s_{A\pi}=s_{AK}~.
\eeq
%%%%%%%%%%%%%%%%%%%%%%%%%%%%%%
\section {Phenomenology of elastic $\pi\pi\to\pi\pi$ scattering} 
\begin{figure}[hbt]
\begin{center}
\includegraphics[width=8.cm]{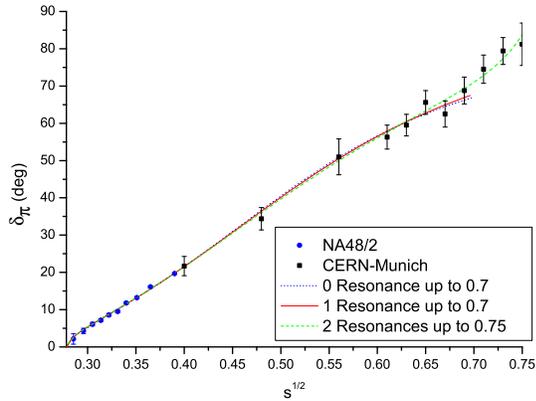}
\vspace*{-0.7cm}
\caption{\footnotesize  Central values of the best fits of the $\pi\pi$ phase shift below $\sqrt{s}$=0.75 GeV for elastic $\pi\pi\to\pi\pi$ scattering: blue (dotted) line for 0 ``bare" resonance with $\chi^2_{min}$ reached at $\sqrt{s}$= 0.7 GeV; red (continuous) line for 1 ``bare" resonance and $\chi^2_{min}$ reached at $\sqrt{s}$= 0.7 GeV; green (dashed) line for 2  ``bare" resonances and $\chi^2_{min}$ reached at $\sqrt{s}$= 0.75 GeV.}
\label{fig:elastic}
\end{center}
\vspace*{-1cm}
\end{figure}

\vspace*{-0.25cm}
\nin
%%%%%%%%%%%%%%%%%%%%%%%%%%
\b {\bf Data input}\\
%%%%%%%%%%%%%%%%%%%%%%%%%%
The only data input used in this process is the pion phase shift $\delta_{\pi}$
well measured experimentally. We shall use the new precise data from NA48/2 on $Ke4\equiv K\to\pi\pi e\nu_e$ for the $\pi\pi$-phase shift below 390 MeV \cite{NA48} and use from 400 to 900 MeV the CERN-Munich \cite{MUNICH} and Hyams et al. \cite{HYAMS} $\pi\pi$-phase from  $\pi\pi\to\pi\pi$ which agree each others above 400 MeV. These data are shown in Fig. \ref{fig:elastic}.
 %%%%%%%%%%%%%%%%%%%%%%%%%%%%%%%
%\\
\nin
 %%%%%%%%%%%%%%%%%%%%%%%%%%%%%%%
 %%%%%%%%%%%%%%%%%%%%%%%%%%%%%%%%%%%%%%%%%%%%%%%%%%
 %\vspace*{-0.5cm}
{\scriptsize
\begin{table}[hbt]
\setlength{\tabcolsep}{0.15pc}
 \caption{\scriptsize    Values in GeV$^{d}~(d=1,2)$ of the bare parameters of the K-matrix model for different ``bare" resonances input for $\pi\pi\to\pi\pi$ elastic scattering. 
 %We have fixed $s_{A_0}=m^2_\pi/2$ at the Adler value. 
 The fit has been performed until $\sqrt{s}\simeq$ 0.7 GeV (0 res. and 1 res.) and 0.75 GeV (2 res.), where the $\chi^2$/ndf  is minimal. The correlated errors of the ``bare" parameters come from the fit of the data using MINUIT. The ones of the physical poles are the quadratic sum of the errors induced by each ``bare" parameters (a linear sum would lead to 2-3 times more accurate values due to cancellations of some of the errors in this case.). An average is given in the last column. }
\begin{tabular}{llllll}
&\\
\hline
%\hline
%\\
Output&0 res. &1res. &2 res.&Average \\
%\\
\hline
%\hline
%\\
$s_{A}$&$0.009(6)$&fixed&fixed&\\
$\sigma_{D\pi}$&$6.2(3.2)$& $1.41(7)$&$1.78(10)$&\\
$\sigma_{D2}$&$7.6\pm 4.5$& $-$&$-$&\\
$s_{R_a}$&$-$&$1.94(9)$&$26.97(1.54)$&\\
$\Lambda$&$108(34)$& $-$&$-$&\\
$g_{\pi a}$&$-$&$2.54(8)$&$10.42(30)$&\\
%\\
$s_{R_b}$&$-$&$-$&$0.61(31)$&\\
$g_{\pi b}$&$-$&$-$&-$0.39(8)$&\\
%&\\
%&\\
${\chi_{min}^2\over {\rm ndf}}$&${12.04\over 14}$=0.86&${11.73\over 15}=0.78$&${12.71\over 16}=0.79$&\\
%&\\
{\small\boldmath$\lrar$}&\\
%&\\
$M_\sigma$&468(181)&456(19) &448(18)&{\it 452(13)}\\
$\Gamma_\sigma/2$&261(211)&265(18) &260(19) &{\it 259(16)}\\
$|g_{\sigma\pi^+\pi^-}|$&2.58(1.31)&2.72(16)&2.58(14) &{\it 2.64(10)}\\
\hline
%\hline
\end{tabular}
\label{tab:elastic}
\end{table}
}
%\\
%\vspace*{-0.5cm}
\nin
%%%%%%%%%%%%%%%%%%%%%%%%%%%%%%%
% \vspace*{-0.75cm}
%%%%%%%%%%%%%%%%%%%%%%%%%%%%%%%%%%%%%%%%%%%%
 %\vspace*{-0.75cm}
{\scriptsize
\begin{table}[hbt]
\setlength{\tabcolsep}{0.2pc}
 \caption{\scriptsize    Mass and 1/2 width in MeV of the $\sigma$ meson in the complex plane. 
    }
\begin{tabular}{lll}
&\\
\hline
%\hline
%\\
Processes&$M_\sigma -i\Gamma_\sigma/2$&Refs.   \\
%\\
\hline
%\hline
\\
This work&\\
$Ke4$ $\oplus$ $\pi\pi\to \pi\pi$&$452(13) -{\rm i}~259(16)$&\\
$Ke4$ $\oplus$ $\pi\pi\to \pi\pi/K\bar K$&$448(43) -{\rm i}~266(43)$&\\
$Average$&${\it 452(12)-{\rm i}~260(15)}$&\\
%&\\
&\\
Others &\\
$\pi\pi\to \pi\pi\oplus$Roy$\oplus$ChPT&$441^{+16}_{-8} -{\rm i}~272^{+9}_{-15}$&\cite{leutwyler}\\
%&\\
$\pi\pi\to \pi\pi/\bar KK\oplus$Roy&$461 \pm 15 -{\rm i} ~(255 \pm 16)$&\cite{GKPY}\\
$J/\psi\to \omega\pi\pi$&$541\pm 39 -{\rm i}~(222\pm 42)$& \cite{BES2}\\
$D^+\to\pi^+\pi^-\pi^+$ &$478 \pm 29-{\rm i}~(162 \pm 46)$&\cite{E741}\\
&\\
\hline
%\hline
\end{tabular}
\label{tab:sigma}
\end{table}
}
%\vspace*{-0.5cm}
\nin
%%%%%%%%%%%%%%%%%%%%%%%%%%%%%%%%%%%%%%%%%%%%%%%
%%%%%%%%%%%%%%%%%%%%%%%%%%
\\
\b {\bf 0 ``bare" resonance $\equiv\lambda \phi^4$ model}\\
%%%%%%%%%%%%%%%%%%%%%%%%%%
Let's first fit the elastic $\pi\pi$ data by using a {\it $\lambda \phi^4$ model without any ``bare" resonance}.
In this old version of the model \cite{MENES}, one can introduce the shape function $f_P$ \cite{MNO}:
\beq
%f_2(s)={s-s_{A2}\over (s + \sigma_{D1})(s + \sigma_{D2})}~,
 T_{PP}= {\Lambda f_P(s) \over 1-\Lambda \tilde f_P(s)},
~~f_P(s)={s-s_{AP}\over (s + \sigma_{D1})(s + \sigma_{D2})}~,
\label{eq:t2f2}
\eeq
where $\sigma_{D1}\equiv \sigma_{D\pi}$ and:
\beq
\tilde f_2(s) = \frac{2}{\pi}  \Big{[} h_2(s)-h_2(0)  \Big{]},
\label{f2tilfct}
\eeq
with :
\begin{eqnarray}
h_P(s) =
f_P(s)\tilde L_{s1}(s)
 - \sum_{i=1,2}\frac{\sigma_{Ni}}{s+\sigma_{Di}}
~\tilde L_{si}(-\sigma_{Di})~.
\label{h2fct}
\end{eqnarray}
$\sigma_{N1},\sigma_{N2}$  in Eq. (\ref{eq:t2f2}) are the residues of $f_P(s)$ at 
$\sigma_{D1}\equiv \sigma_{D\pi}$, $\sigma_{D2}$. 
In fitting the ``bare" parameters, we look for a minimum of $\chi^2$$\equiv \chi^2_{min}$
by varying the range of the interval $[4m_\pi^2,s]$ inside which we perform the fit. Here, this is obtained for $\sqrt{s}$=0.7 GeV where:  $\chi^2_{min}$/ndf=12.04/14=0.86. The fitted values of the ``bare" parameters and the resulting values of the physical pole parameters are given in Table\,\ref{tab:elastic}\,\footnote{Fixing the Adler value at the one in Eq. \ref{eq:adler} does not bring any improvements here.}.
The quoted errors of the ``bare" parameters come from the fit program MINUIT. The errors induced by each of these ``bare" parameters on the physical poles can be added (as currently done) linearly or quadratically \footnote{Notice that if we have added linearly the errors induced by each ``bare" parameters by taking into account their signs, we would have obtained about 2 times more accurate predictions. For a more conservative error, we shall take here and in the following the quadratic sum of these errors.}.
These results indicate that, though not accurate, this
original version of the model gives a reasonnable value of the physical parameters.
%%%%%%%%%%%%%%%%%%%%%%%%%%%
\\
\b {\it \bf %1~channel $\oplus$ 
1~``bare" resonance}\\
%%%%%%%%%%%%%%%%%%%%%%%%%%%
This analysis has been done in \cite{MNO} using the CGL parametrization 
based on Roy equations with constraints from chiral symmetry \cite{leutwyler}.
In the following, we shall use instead  the new precise data from NA48/2 on $Ke4$ for the $\pi\pi$-phase below 390 MeV \cite{NA48} and use from 400 to 900 MeV the CERN-Munich \cite{MUNICH} and Hyams et al. \cite{HYAMS} $\pi\pi$-phase from  $\pi\pi\to\pi\pi$ which agree each others above 400 MeV. We  extract the ``bare" parameters  from these data:\\
-- In the first step, we leave all ``bare" parameters free and find a minimum $\chi^2$: $\chi_{min}^2$/ndf=9.43/17=0.55 for $\sqrt{s}=0.75$ GeV. The fitted value of the Adler zero is:
\beq
s_{A\pi}=0.0394(92)~{\rm GeV^2}~,
\eeq
which is relatively bad compared with the theoretical expectation\,\footnote{Here and in the following, we shall not explictly include isospin breaking effects which we expect to give small corrections.} 
:
\beq
s_{A\pi}={m_{\pi}^2\over 2}=0.0094~{\rm GeV^2}~.
\label{eq:adler}
\eeq
-- Then in the second step, we fix the Adler zero at the value in Eq. \ref{eq:adler} and deduce the results
in Table\,\ref{tab:elastic}.
The fit is shown in Fig. \ref{fig:elastic}. 
These
``bare" parameters lead to the physical poles  in Table \ref{tab:elastic},
which we consider as improvements of the previous results in \cite{MNO}.  
This result is comparable in size and errors with the precise determinations from recent analyses of the analogous $\pi\pi\to\pi\pi/\bar KK$ scatterings data using different approaches (Roy equations $\oplus$ chiral symmetry constraints \cite{leutwyler}, Roy equations $\oplus$ control of the high-energy behaviour of the amplitude \cite{GKPY})(Table \ref{tab:sigma})\,\footnote{See also Table \ref{tab:sigma} for some other determinations.}, which have been obtained before the last $Ke4$ NA48/2 precise data \cite{NA48}. 
%%%%%%%%%%%%%%%%%%%%%%%%%%%
\\
\b {\it \bf %1~channel $\oplus$ 
2~``bare" resonances}\\
%%%%%%%%%%%%%%%%%%%%%%%%%%%
We repeat the previous analysis by working instead with 2~``bare" resonances. We fix the Adler zero at the value in Eq. \ref{eq:adler} and fit the other ``bare" parameters. We obtain the results quoted in Table \ref{tab:elastic} for a $\chi_{min}^2$/ndf=12.71/16=0.794 at $\sqrt{s}=0.75$ GeV. The fit is shown in Fig. \ref{fig:elastic}. 
%%%%%%%%%%%%%%%%%%%%%%%%%%
\\
\b {\bf Comments and final results from  $\pi\pi\to\pi\pi$}
\\
%%%%%%%%%%%%%%%%%%%%%%%%%%
From previous studies, we conclude that: \\
--  The results from different forms of the model in Table \ref{tab:elastic} are very stable. The final results from  elastic $\pi\pi\to\pi\pi$ are the average of the ones from 0, 1 and 2 ``bare" input resonances quoted in this Table  \ref{tab:elastic}, which are:
\bea
M_\sigma[{\rm MeV}]&=& 452(13)-{\rm i}~259(16)~, \nnb\\
|g_{\sigma\pi^+\pi^-}|&=&2.64(10)~{\rm GeV}~.
\label{eq:elasticfin}
\eea
-- The results, from the 0 ``bare" resonance or $\lambda\phi^4$ model 
show that the existence of the $\sigma$ pole is not an artifact of the ``bare" resonance
entering in the parametrization of the $\pi\pi$ amplitude $T_{PP}$.\\
Noting that the concavity of the fit curve in Fig. \ref{fig:elastic} around the $\rho$-meson mass region has raised some doubts on the data of the phase shift $\delta_\pi$ \cite{MINKOWSKI}, we have redone the fit by assuming that the data increases linearly from the $Ke4$ one. Using 1 or 2 resonances, we still find, in this extreme case, a $\sigma$ pole :
\beq
M_\sigma[{\rm MeV}]\simeq 413-{\rm i}~300~,
\eeq
where a similar value has been obtained earlier \cite{MINKOWSKI}. This result may
indicate that the existence and the dynamics of the $\sigma$ is mainly due to the low-energy 
behaviour of the $\pi\pi$ phase shift $\delta_\pi$ data, which are accurately determined from $Ke4$ by NA48/2 \cite{NA48}.
%%%%%%%%%%%%%%%%%%%%%%%%%%%%%%%
%\vspace*{-0.5cm}
\section{Phenomenology of inelastic $\pi\pi\to\pi\pi/\bar KK$}
\vspace*{-0.25cm}
\nin
%%%%%%%%%%%%%%%%%%%%%%%%%%%%%%%
%\section{Phenomenology of $\pi\pi\to\pi\pi/\bar KK$}
%\vspace*{-0.25cm}
%\nin
 %%%%%%%%%%%%%%%%%%%%%%%%%
 %%%%%%%%%%%%%%%%%%%%%%%%%%
\b {\bf 2 ``bare" resonances $\oplus$ 2 channels   parameters}\\
%%%%%%%%%%%%%%%%%%%%%%%%%%
 %%%%%%%%%%%%%%%%%%%%%%%%%%%%%%%
\begin{figure}[hbt]
\begin{center}
\includegraphics[width=7.cm]{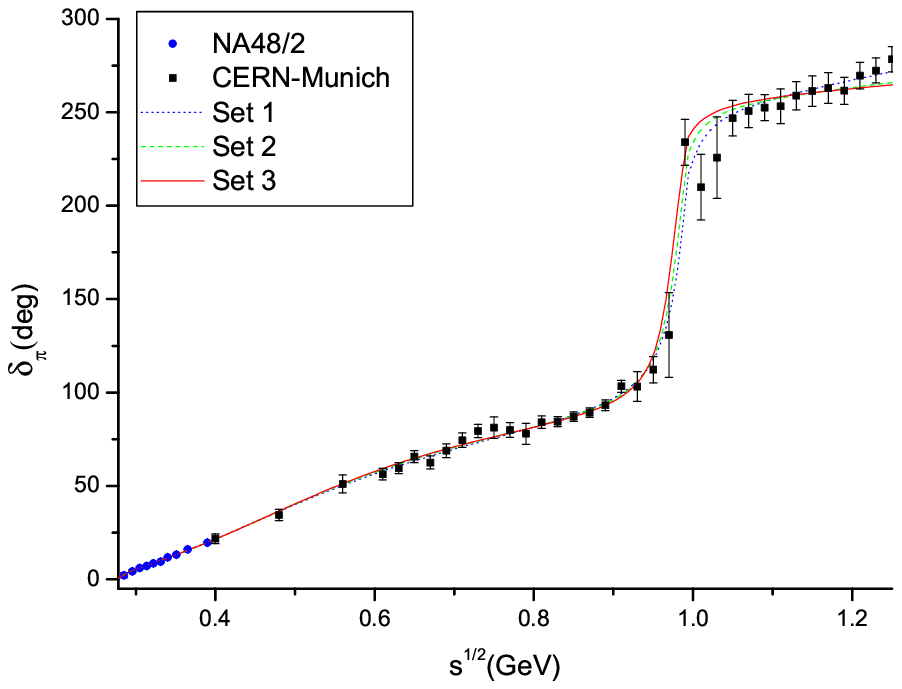}
\includegraphics[width=7.cm]{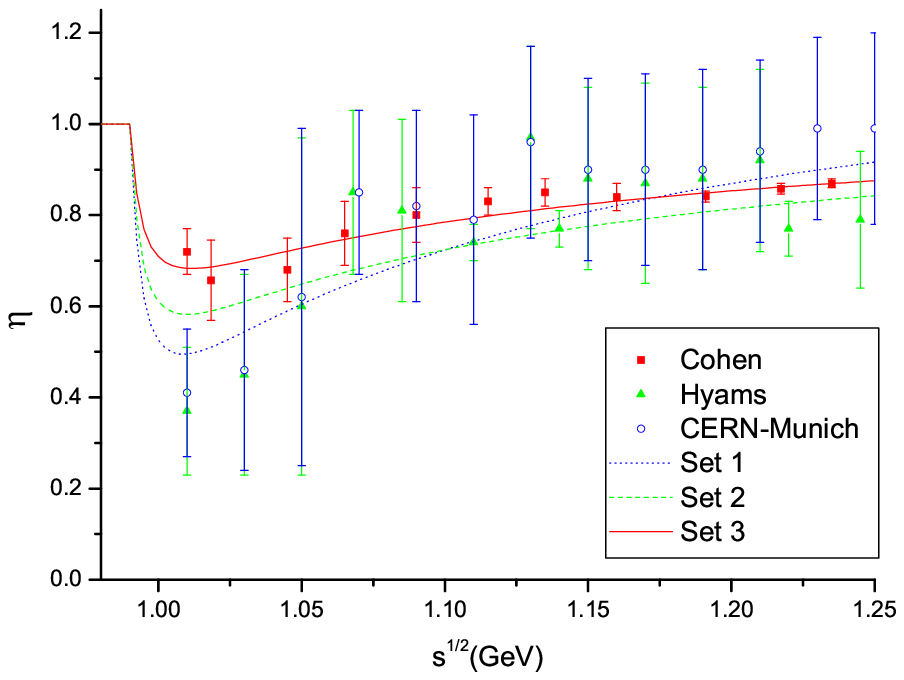}
\includegraphics[width=7.cm]{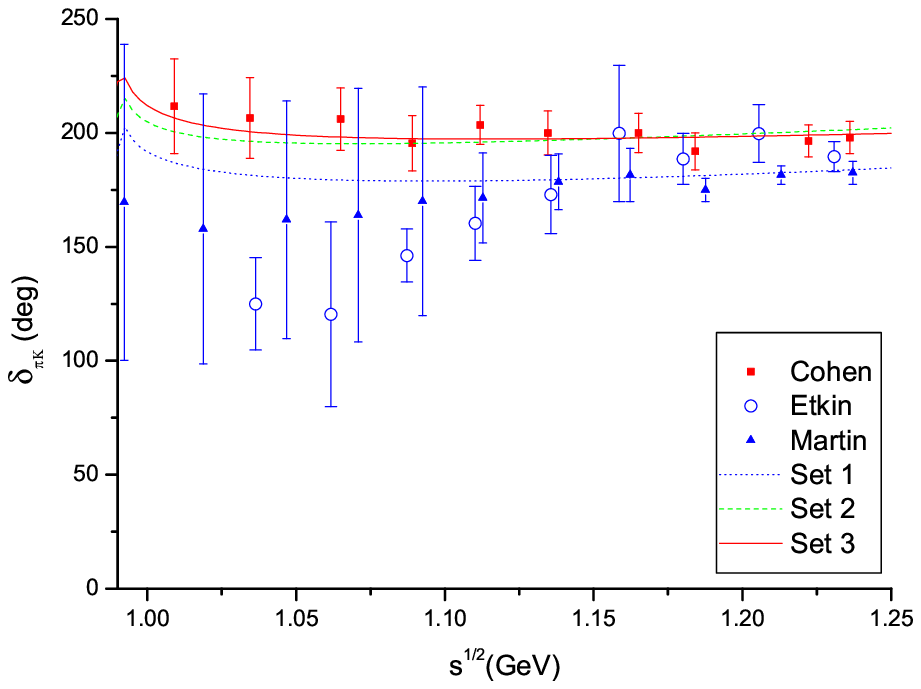}
\vspace*{-0.7cm}
\caption{\footnotesize  a) Fit of the $\pi\pi$ phase $\delta_\pi$ versus $\sqrt{s}$: Set 1 (blue: dotted); Set 2 (green: dashed); Set 3 (red: continuous);
b) Fit of the inelasticity $\eta$; c) Fit of the sum of $\pi$ and $K$ phase $\delta_{\pi K}.$}
\label{fig:fit}
\end{center}
\vspace*{-1.cm}
\end{figure} 
%\\
 %%%%%%%%%%%%%%%%%%%%%%%%%%%%%%%
In so doing, we take in Table\,\ref{tab:set} three representatives sets of  $\pi\pi\to\pi\pi/\bar KK$ data in the existing literature:
%%%%%%%%%%%%%%%%%%%%%%%%%%%%%%%
% \vspace*{-0.5cm}
{\scriptsize
\begin{table}[hbt]
%\bea
\setlength{\tabcolsep}{0.9pc}
 \caption{\scriptsize  Different data used for each different {\it sets} for determining the ``bare" parameters
 in Table \ref{tab:bare}: $\delta_\pi$is the $\pi\pi$ phase shift, $\eta$ is the inelasticity and $\delta_{\pi K}$ is the sum of the $\pi$ and $K$ phases.}
\begin{tabular}{lllll}
&\\
\hline
%\hline
%\\
Input&Set 1&Set 2 &Set 3  \\
%\\
\hline
%\hline
%\\
$\delta_\pi$&\cite{NA48,MUNICH,HYAMS}&\cite{NA48,MUNICH,HYAMS}&\cite{NA48,MUNICH,HYAMS}\\
$\eta$&\cite{MUNICH}&\cite{HYAMS}&\cite{COHEN}\\
$\delta_{\pi K}$&\cite{ETKIN}&\cite{COHEN}&\cite{COHEN}\\
%&\\
\hline
%\hline
\end{tabular}
\label{tab:set}
\end{table}
%\eea
}
%\\
%\vspace*{-0.5cm}
\nin
%%%%%%%%%%%%%%%%%%%%%%%%%%%%%%%%%%%%%%
 %%%%%%%%%%%%%%%%%%%%%%%%%%%%%%%
\begin{figure}[hbt]
\begin{center}
\includegraphics[width=7.cm]{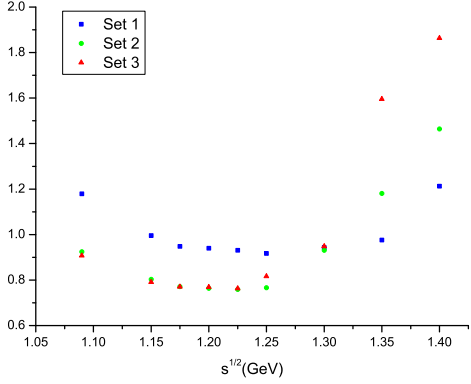}
\vspace*{-0.7cm}
\caption{\footnotesize  $\chi^2$/ndf versus $\sqrt{s}$ in GeV : Set 1 (blue: square); Set 2 (green: circle); Set 3 (red: triangle).}
\label{fig:chi2}
\end{center}
\vspace*{-1.cm}
\end{figure} 
\\
\nin
 %%%%%%%%%%%%%%%%%%%%%%%%%%%%%%%
 -- The choice for $\delta_\pi$ which we have used in the analysis of elastic $\pi\pi\to\pi\pi$ scattering  
 is unique and comes from the new $Ke4$  data of NA48/2 \cite{NA48} below 390 MeV and from  $\pi\pi\to\pi\pi/\bar KK$  data above 400 MeV measured by CERN-Munich \cite{MUNICH} and Hyams et al. \cite{HYAMS}  [see Figs. \ref{fig:elastic} and \ref{fig:fit} a)]. \\
-- For the inelasticity $\eta$, different data exhibits a minimum $\eta_{min}$ just above the $\bar KK$ threshold.  CERN-Munich and Hyams et al.  give the smallest value: $\eta_{min}\approx 0.4$, while
Cohen et al. \cite{COHEN} provide the largest one:  $\eta_{min}\approx 0.7$ [see Fig. \ref{fig:fit} b)]. \\
-- For the sum of $\pi$ and $K$ phase $\delta_{\pi K}$, we use the one from Cohen et al and from Etkin-Martin \cite{ETKIN}, which represent the two extreme cases [see Fig. \ref{fig:fit} c)].\\
With these choices, we expect to span all possible regions of the space of parameters, and then to
extract results which do not only come from a single experiment. We have not used the data of Kaminski et al. \cite{KLL} due to the large errors, which, however, agree within the errors with the other data {\it sets} used here.  \\
{\scriptsize
\begin{table}[hbt]
\setlength{\tabcolsep}{0.25pc}
 \caption{\scriptsize    Values in GeV$^{d}~(d=1,2)$ of the bare parameters of the K-matrix model for 2 channels $\oplus$ 2 bare resonances from $Ke4~\oplus~\pi\pi\to\pi\pi/\bar KK$ scatterings. 
  The fit has been performed until $\sqrt{s}\simeq 1.225\sim 1.250$ GeV, where the $\chi^2$/ndf  is minimal (see Fig. \ref{fig:chi2}). The correlated errors come from the fiitting procedure
 using the program MINUIT.}
\begin{tabular}{lllll}
&\\
\hline
%\hline
%\\
Output&Set 1&Set 2 &Set 3  \\
%\\
\hline
%\hline
%\\
$s_{A}$&$0.016\pm 0.004$&$0.013\pm 0.006$&$0.010\pm 0.006$\\
$\sigma_{D}$&$0.740\pm 0.097$& $0.909\pm 0.201$&$1.116\pm 0.262$\\
$s_{R_a}$&$4.112\pm 0.499$&$2.230\pm 0.271$&$2.447\pm 0.298$\\
$g_{\pi a}$&-$0.557\mp 0.177$&$0.846\pm 0.391$&$0.997\pm 0.516$\\
$g_{K a}$&$3.191\pm0.499$&$1.458\pm 0.262$&$1.684\pm 0.363$\\
\\
$s_{R_b}$&$1.291\pm 0.062$&$1.187\pm 0.094$&$1.354\pm 0.149$\\
$g_{\pi b}$&-$1.562\mp 0.117$&-$1.527\mp 0.134$&-$1.756\mp 0.183$\\
$g_{Kb}$&$0.748\pm 0.062$&$0.999\pm 0.149$&$1.159\pm 0.261$\\
&\\
${\chi_{min}^2\over {\rm ndf}}$&${70.6\over 77}$=0.914&${48.8\over 64}=0.759$&${44.3\over 58}=0.763$\\
%&\\
\hline
%\hline
\end{tabular}
\label{tab:bare}
\end{table}
}
%\\
%\vspace*{-0.5cm}
\nin
%%%%%%%%%%%%%%%%%%%%%%%%%%%%%%%
% \vspace*{-0.75cm}
%{\scriptsize
\begin{table}[hbt]
\setlength{\tabcolsep}{0.4pc}
 \caption{\scriptsize    $\sigma$ and $f_0(980)$ meson parameters from $\pi\pi\to\pi\pi/\bar KK$ scatterings using the bare parameters in Table \ref{tab:bare} : the mass and width are in MeV, while the couplings are in GeV.  The errors are the quadratic sum of the ones induced by the ``bare" parameters in Table \ref{tab:bare} (a linear sum would lead to 2-3 times more accurate values due to cancellations of some of the errors in this case). 
An average is given in the last column. }
\begin{tabular}{lllll}
&\\
\hline
%\hline
%\\
Outputs&Set 1&Set 2 &Set 3 & {\it Average}   \\
%\\
\hline
%\hline
%\\
$M_\sigma$&435(74)&452(72) &457(76)&{\it 448(43)}\\
$\Gamma_\sigma/2$&271(92)&266(65) &263(72) &{\it 266(43)}\\
$|g_{\sigma\pi^+\pi^-}|$&2.72(78)&2.74(61)&2.73(61) &{\it 2.73(38)}\\
$|g_{\sigma K^+K^-}|$&1.83(86)&0.80(55)&0.99(68) &{\it 1.06(38)}\\
%$r_{\sigma\pi K}$\\
\\
$M_f$&989(80)&982(47)&976(60)&{\it 981(34)}\\
$\Gamma_f/2$&20(32)&18(16)&18(18)&{\it 18(11)}\\
$|g_{f\pi^+\pi^-}|$&1.33(72)&1.22(60)&1.12(31)&{\it 1.17(26)}\\
$|g_{f K^+K^-}|$&3.21(1.70)&2.98(70)&3.06(1.07)&{\it 3.03(55)}\\
%$r_{f\pi K}$\\
%&\\
\hline
%\hline
\end{tabular}
\label{tab:param}
\end{table}
 \vspace*{-.25cm}
%}
\nin
\\
%%%%%%%%%%%%%%%%%%%%%%%%%%%%%%%
 %%%%%%%%%%%%%%%%%%%%%%%%%%%%%%%
In the following, we shall use:
\beq
m_K\equiv {1\over 2}\ga {m_{K^+}+m_{K^0}}\dr = 495.65 ~{\rm MeV}~.
\eeq
Letting all ``bare" parameters free, we study in Fig.\,\ref{fig:chi2}, using the fitting program MINUIT, the variation of $\chi^2$/ndf versus $\sqrt{s}$ until 1.4 GeV where the data are available. In the fitting procedure, we have chosen the same initial conditions for the 3 {\it sets}, where a good convergence with a good $\chi^2$/ndf of the solutions has been obtained for Set 2 and Set 3. 
A minimum value $\chi^2_{min}$/ndf is reached for $\sqrt{s}\simeq (1.225-1.250) $ GeV at which we extract the optimal outputs given in Table \ref{tab:bare}.  
 At each corresponding value of $\chi^2_{min}$/ndf, the fits for different {\it sets} of data are shown in 
 Fig.\,\ref{fig:fit}. All three {\it sets} give good values of  $\chi^2_{min}$/ndf less than one. 
 %%%%%%%%%%%%%%%%%%%%%%%%%
 \\
\b {\bf Poles  from 2 ``bare" resonances $\oplus$ 2 channels }\\
%%%%%%%%%%%%%%%%%%%%%%%%%
We use the results of the ``bare" parameters  in Table\,\ref{tab:bare} obtained at $\chi^2_{min}$/ndf for deducing the ones of the complex poles in Table\,\ref{tab:param}. The errors on the physical poles are induced by the ones of the ``bare" parameters  in Table\,\ref{tab:bare} and have been added quadratically. 
The iteration of solutions from Set 1 has only a local minimum in $\chi^2$ such that, in order to be more conservative, we have multiplied by a factor 2 the related uncertainties of the results coming from the fit. 
The last column gives the mean value from the three different determinations. We have taken (as is usual in the literature) the weighted average, where the corresponding error is more weighted by the most accurate predictions\,\footnote{Alternatively, we can take, as a final value, the most accurate prediction, which leads about the same result.}. One can see in Table\,\ref{tab:param} that the results from different sets of data are {\it unexpectedly} stable for both $\sigma$ and $f_0(980)$ parameters, which increase our confidence on their independence on the input data {\it sets}.  \\
\nin
%%%%%%%%%%%
-- For the $\sigma$, we obtain the average of the complex pole  mass and width given in Table \ref{tab:param}:
\beq
M_\sigma[{\rm MeV}]=448(43)-{\rm i}~266(43)~.
\label{eq:inelastic}
\eeq
This result is in perfect agreement with the mean value from elastic $\pi\pi\to\pi\pi$ scattering in Eq. \ref{eq:elasticfin} and comparable in size and errors with the ones in Table \ref{tab:sigma}. Averaging the
two predictions in Eqs. \ref{eq:elasticfin} and \ref{eq:inelastic}, we deduce our final value:
\beq
M_\sigma[{\rm MeV}]=452(12)-{\rm i}~260(15)~.
\label{eq:msig}
\eeq
Averaging the result in Table \ref{tab:param} and Eq. \ref{eq:elasticfin}, we deduce:
\bea
|g_{\sigma\pi^+\pi^-}|&=&2.65(10)~{\rm GeV},\nnb\\
r_{\sigma\pi K}&\equiv& {|g_{\sigma K^+K^-}|\over |g_{\sigma\pi^+\pi^-}|}=0.37(6)~,
\label{eq:rspik}
\eea
which improves and confirms our previous rough findings in \cite{KMN} and which is comparable with some other determinations in Table\,\ref{tab:coupling} from \cite{KMN}\,\footnote{Similar values of $r_{\sigma\pi K}$   are also found from some fits  in \cite{ACHA} but the results obtained there are unstable.}: {\it the sizeable coupling of the $\sigma$ to $\bar KK$
disfavours the usual  $\pi\pi$ molecule and four-quark assignement of the $\sigma$, where this coupling
is expected to be negligible.}\\
-- For the $f_0(980)$, we obtain the mean value:
\beq
M_f[{\rm MeV}]=981(34)-{\rm i}~18(11)~,
\label{eq:mf}
\eeq
which is comparable with the PDG range of values\,\cite{PDG}:
\beq
M_f[{\rm MeV}]=980(10)-{\rm i}(20\sim 50)~.
\eeq
From Table \ref{tab:param}, we also find:
\bea
|g_{f\pi^+\pi^-}|&=&1.12(31)~{\rm GeV},\nnb\\
r_{f\pi K}&\equiv& {|g_{fK^+K^-}|\over |g_{f\pi^+\pi^-}|}=2.59(1.34)~,
\label{eq:rfpik}
\eea
in agreement with the determinations in the existing literature (see Table\,\ref{tab:coupling}). {\it The large value of this ratio of coupling and the relative narrowness of the $f_0(980)$ width (compared to e.g. the $\rho$-meson)  does not favour the pure ($\bar uu+\bar dd$) content of the $f_0(980)$ where $r_{f\pi K}$ is expected to be about 1/2 and the width of about 120 MeV \cite{SNG,SN06}. This feature has been used as an indication of the four-quark nature of the $f_0(980)$ (see e.g. \cite{ACHA}) or alternatively of its large gluonium component via a maximal mixing with a $\bar qq$ state (see e.g. \cite{BN,SNG})}. 
%%%%%%%%%%%%%%%%%%%%%%%%%%%%%%%
%\vspace*{-0.3cm}
{\scriptsize
\begin{table}[hbt]
\setlength{\tabcolsep}{0.0pc}
 \caption{\scriptsize    Modulus of the $\pi^+\pi^-$ and $K^+K^-$ complex couplings in GeV of the $\sigma $ and of $f_0(980)$ from S- and K-matrix models for $\pi\pi\to \pi\pi/K\bar K$ scatterings compared with the ones from $\phi$ and $J/\psi$ decays. $r_{S\pi K}\equiv  |g_{S K^+K^-}|/|g_{S\pi^+\pi^-}|$: $S\equiv \sigma,~f$. 
    }
\begin{tabular}{lccccc}
&\\
\hline
 Processes&$\vert g_{\sigma\pi^+\pi^-}|$ & $r_{\sigma\pi K}$ &$|g_{f\pi^+\pi^-}|$ &$r_{f\pi K}$&Models  \\
%\\
\hline
This work &\\
%\\
$\pi\pi\to \pi\pi/K\bar K$&$2.65(10)$&$0.37(6)$&1.17(26)&2.6(1.3)&\cite{MNO,MENES} \\
&\\
Others &\\
$\pi\pi\to \pi\pi/K\bar K$ &$2.03(3)$&0.65(18)&0.97(6)&1.7(2)&\cite{KLL} \\
%\\
&2.5&0.62&1.55&1.20&\cite{WANG}\\
%&\\
$\phi\to \sigma /f_0(980)~\gamma$&$-$&0.67&$-$&$-$& \cite{BUGG}\\
%&\\
$\psi\to\phi~ \pipi/K\bar K$&$-$&$-$&2.35&1.80& \cite{BES}\\
%&\\
%\hline
&\\
{\it Average}&{\it 2.4}&{\it 0.6}&{\it 1.5}&{\it 1.8}& \\
%&\\
\hline
%\hline
\end{tabular}
\label{tab:coupling}
\end{table}
}
%\vspace*{-0.5cm}
\nin
\\
%%%%%%%%%%%%%%%%%%%%%%%%%%
\b {\bf Model with 1 ``bare" resonance $\oplus$ 2 channels}\\
%%%%%%%%%%%%%%%%%%%%%%%%%%
We have further studied the influence of some other configurations of the model on the fit of the $\sigma$
by analyzing the {\it minimal} case: {\it 1 ``bare" resonance $\oplus$ 2 channels} and
using for instance Set 3 of data.  Letting all ``bare" parameters free, the $\chi^2_{min}$/ndf $\simeq 10$  is very bad which is obtained by doing the fit from $2m_\pi$ to $\sqrt{s}=1.2$ GeV. As (intuitively) expected, the previous results for the $\sigma$ parameters are approximately reproduced:
\beq
M_\sigma[{\rm MeV}]\approx 377-{\rm i}~195~, 
\eeq
and
\beq
|g_{f\pi^+\pi^-}|\approx 2.13~{\rm GeV},~~~r_{\sigma\pi K}\approx 0.42~,
\eeq 
while the $f_0$ mass is pushed far away from the $\bar KK$ threshold: 
\beq
M_f[{\rm GeV}]\approx 3.8+{\rm i}~1.7~.
\eeq
Due to the bad quality of $\chi^2$/ndf, the result from this version of the model will not be retained. 
%%%%%%%%%%%%%%%%%%%%%%%%%%%%%%
\section {On-shell mass, width and couplings of the $\sigma$} 
\vspace*{-0.25cm}
\nin
%%%%%%%%%%%%%%%%%%%%%%%%%%%%%%
Due to the large width of the $\sigma$, a direct comparison of the previous results with the ones obtained from
QSSR or some other theoretical predictions in the real axis is questionable. For better comparing the results obtained in the complex plane with the theoretical predictions obtained in the real axis, it is more appropriate to introduce like in \cite{MNO} the on-shell meson \cite{SIRLIN} masses and hadronic widths, where the amplitude is purely imaginary 
at the phase 90$^0$: 
\beq
{\rm Re} {\cal D}({(M^{\rm os}_\sigma})^2)=0\lrar M_\sigma^{\rm os}\simeq 0.9~{\rm GeV}~. 
\eeq
In the same way as for the mass, one can also define an ``on-shell width" \cite{MNO}
 from  Eqs. (\ref{eq:imaginary}) and (\ref{eq:gpi2}) evaluated at $s=(M^{\rm os}_\sigma)^2$ :
\beq
M^{\rm os}_\sigma \Gamma^{\rm os}_\sigma\simeq {{\rm Im}~ \calD\over {\rm Re~} {\cal D}'}
\lrar \Gamma_{\sigma\to\pi^+\pi^-}^{\rm os}\simeq 0.7~{\rm GeV}~,
\eeq
which are comparable with the Breit-Wigner mass and width \cite{HYAMS,ESTABROOKS,AMP}:
\beq 
M_{BW}\simeq \Gamma_{BW}\simeq 1~{\rm GeV}~.
 \label{sigmamass} 
\eeq 
These values lead to the on-shell coupling:
\beq
\vert g_{\sigma\pi^+\pi^-}^{\rm os}\vert \simeq 6~{\rm GeV}~.
\label{sigmacoupl}
\eeq
%%%%%%%%%%%%%%%%%%%%%%%%%%%%%%%
\section{Comparison with QSSR $\oplus$ LET predictions}
\vspace*{-0.25cm}
\nin
%%%%%%%%%%%%%%%%%%%%%%%%%%%%%%%
-- One on hand, the corresponding on-shell (or Breit-Wigner) mass and coupling of the $\sigma$ are comparable in size with the predictions from combined QSSR $\oplus$ LET analysis \cite{SNG,SNG0,VENEZIA} for a glueball with a large OZI violation for its coupling to $\pi\pi$ and $\bar KK$\,\footnote{The expectation of  a glueball chiral coupling to pair of Goldstone bosons \cite{CHANOW} could not hold in this non-perturbative regime.} :
\beq
M_{\sigma}\simeq 1~{\rm GeV}~,~~~~~\vert g_{\sigma\pi^+\pi^-}\vert \simeq \vert g_{\sigma K^+K^-}\vert \simeq5~{\rm GeV}~,
\eeq
implying:
\beq
\Gamma_{\sigma\to\pi^+\pi^-}={\vert g_{\sigma\pi^+\pi^-}\vert ^2\over 16\pi M_{\sigma}}\sqrt{1-{4m_\pi^2\over M_\sigma^2}}\simeq 0.7~{\rm GeV}~.
\eeq
The existence of the $\sigma$ is necessary for a consistency between the subtracted and unsubtracted QSSR~\cite{VENEZIA},
where the gluonium two-point correlator subtraction constant\,\cite{NOVIKOV}:
\beq
\psi(0)\simeq  -{1\over 16}{\beta_1\over\pi}\la\alpha_s G^2\ra~
\eeq
plays a crucial role ($\beta_1=-11/2+n/3$ for $n$ flavours), and where  the value of the gluon condensate is:  $\la\alpha_s G^2\ra=(6.8\pm 1.3)\times 10^{-2}$ GeV$^4$\,\cite{SNHeavy,LNT,SNI,SNTAU}.\\
-- On the other hand, QSSR predicts for a $S_2\equiv 1/\sqrt{2}(\bar uu+\bar dd)$  I=0 scalar meson  \cite{SNG,SN06}:
\beq
M_{S_2} \simeq 1~{\rm GeV}~,~~~\Gamma_{S_2}\simeq 0.12~{\rm GeV}~,
\label{eq:s2mass}
\eeq
and 
\beq
~~\vert g_{S_2\pi^+\pi^-}\vert \simeq 2.5~{\rm GeV}~,~~~{\vert g_{S_2K^+K^-}\vert\over \vert g_{S_2\pi^+\pi^-}\vert}={1\over 2}~.
\label{eq:s2width}
\eeq
These results indicate that:\\
--The $S_2$ is narrower and much higher in mass than the complex $\sigma$ pole often identified with a $\bar qq$ state in the current literature. \\
-- The $f_0(980)$ cannot be a pure $(\bar uu+\bar dd)$ state due to the large ratio of its  $\bar KK$ over its $\bar \pi\pi$ couplings $r_{f\pi K}$ [Eq.\,\ref{eq:rfpik}] and to its relative (compared to the $\rho$-meson) small $\pi\pi$ width.  It cannot be also a pure $\bar ss$ or $\bar KK$ molecule due to its non-negligible width into $\pi\pi$. \\
-- A large gluonium component eventually mixed with a $\bar qq$ state in the $\sigma$ and $f_0(980)$  \cite{SN09,BN,SNG,MINK,VENEZIA,MNO,KMN} can be advocated for evading these  previous difficulties. \\
-- Further phenomenological searches for gluonium have been proposed in the literature in $\phi,J/\psi$ and $\Upsilon$ radiative decays\,\cite{SNG,SN06,VENEZIA}, $D,~B$ semi-leptonic \cite{DOSCH,SN06} and hadronic \cite{MINK} decays. 
%%%%%%%%%%%%%%%%%%%%%%%%%%%
%\vspace*{-.15cm}
\section{Summary and conclusions}
\vspace*{-0.25cm}
\nin
%%%%%%%%%%%%%%%%%%%%%%%%%%%
 We have used new $Ke4\equiv K\to\pi\pi e\nu_e$ on $\pi\pi$ phase shift below 390 MeV $\oplus$ different $\pi\pi\to \pi\pi / K\bar K$ scatterings data above 400 MeV, for extracting the $\sigma\equiv f_0(600)$ and $f_0(980)$ masses, widths and hadronic couplings, within an improved ``analytic $K$-matrix model". \\
 \b Using a $\lambda \phi^4$ version of the model, we have noticed from different analysis that the existence of the $\sigma$ in the complex plane and having a mass of about 452 MeV  is not an artifact of ``bare" resonances used in the analytic K-matrix model. \\
\b We have also seen that our predictions are very stable vesus the different forms (number of ``bare" resonances) of the models. Our results are summarized in Table\,\ref{tab:param}. \\
\b The masses and widths [Table \ref{tab:sigma} and Eqs. \ref{eq:msig}] of the $\sigma$
are comparable in size and errors with the most accurate determinations  in the existing literature \cite{leutwyler,GKPY} (see also Table \ref{tab:sigma}), while the ones of the $f_0(980)$ in Eq.  \ref{eq:mf} are comparable with the PDG values \cite{PDG}. The small uncertainties in our determinations can be mainly due to the new accurate data on Ke4 from NA48/2.\\
\b The values of the couplings confirm and improve our previous results in \cite{KMN} and are comparable with the ones from some other processes given in Table \ref{tab:coupling}: \\
-- The (unexpected) sizeable coupling of the $\sigma$ to $\bar KK$: $r_{\sigma\pi K}\simeq 0.37(6)$ [Eq. \ref{eq:rspik}] is a strong indication against a pure $\bar \pi\pi$ molecule and four-quark substructure of the $\sigma$, whilst its large width cannot be explained (using the QSSR results in a previous section) from a simple ($\bar uu+\bar dd$) assignement.  \\
-- The large value: $r_{f\pi K}\simeq 2.59(1.34)$ [Eq. \ref{eq:rfpik}]
of the ratio of the $f_0(980)$ couplings to $\bar KK$ over the one to $\bar \pi\pi$ and of the $f_0(980)$ relative narrow width compared e.g. with the one of the $\rho$-meson does not favour the pure ($\bar uu+\bar dd$) assignement of the $f_0(980)$. In this scheme one would predict a ratio of coupling of about 1/2 and a width of about 120 MeV [Eq. \ref{eq:s2mass}]. \\
\b The four-quark scenario can explain the large $\bar KK$ coupling of the $f_0(980)$ but it fails to explain the large coupling of the $\sigma$ to $\bar KK$.  \\
\b The simple $\bar qq$ scheme cannot explain the large $\bar KK$ coupling and the narrowness of the $f_0(980)$ as well as the broad width of the $\sigma$. \\
\b A large gluonium component eventually mixed with a $\bar qq$ state of the $\sigma$ and $f_0(980)$ can be advocated  \cite{SN09,BN,SNG,MINK,VENEZIA,MNO,KMN} for evading these above-mentioned difficulties.  
  %%%%%%%%%%%%%%%
%\vspace*{-0.35cm}
\section*{Acknowledgements} 
\vspace*{-0.25cm}
\nin
We thank Robert Kaminski, Peter Minkowski and Wolfgang Ochs for some email exchanges and for some comments. 
G.M and S.N. acknowlege the IN2P3-CNRS for a partial support within the project ``Non-perturbative QCD and Hadron Physics".  
\vfill\eject
%\input{bib_sigmab}
%%%%%%%%%%%
%%%%%%%%%%%%%%%

\end{document}